\documentclass[aps,amsmath,amssymb,prl,twocolumn,showpacs,nofootinbib]{revtex4-1}
\usepackage{bm}
\usepackage{epsfig}
\usepackage[usenames]{color}
\newcommand{\eqa}{\begin{eqnarray}}

\newcommand{\eqe}{\end{eqnarray}}

\newcommand{\lk}{\left(}    \newcommand{\rk}{\right)}

\newcommand{\lK}{\left[} \newcommand{\rK}{\right]}

\newcommand{\bnabla}{\mbox{\boldmath$\nabla$}}

\newcommand{\cF}{\mbox{$\cal F$}}

\newcommand{\bp}
{\mbox{\boldmath$p$}}

\newcommand{\br}{\mathbf r}%
\newcommand{\bu}{\mbox{\boldmath${u}$}}

\newcommand{\bomega}{\mbox{\boldmath$\omega$}}

\newcommand{\phibar}{\phi\kern-1ex\rule[1.25ex]{1ex}{.1ex}}

\begin{document}
\setlength{\parindent}{0cm}

\title{Pre-melting of crossing vortex lattices}

\author{S. Butsch$^1$, E. Zeldov$^2$ and T. Nattermann$^3$}\affiliation{$^1$Miamed GmbH, K\"oln, Germany} 
\affiliation{$^2$Weizmann Institute of Science, Rehovot 76100, Israel}\affiliation{$^3$Institut f\"ur Theoretische Physik, Universit\"at zu K\"oln,  K\"oln, Germany}\date{\today}
\begin{abstract}
The pre-melting of high vortex density planes observed recently in layered superconductors in tilted magnetic field \cite{Segev+11} is explained theoretically. Based on the structural information of the crossing lattices of pancake and Josephson vortices the effective vortex cage potential at different lattice sites is determined numerically. Melting takes place when the thermal energy allows proliferation of vacancy-interstitial pairs. It is found that the increased density of pancake vortex stacks in the planes containing Josephson vortices, rather than their incommensurate structure, is the main cause for pre-melting.
\end{abstract}

{}

\maketitle

Melting of solids, an everyday phenomenon from the shrinking ice cube in a drink to the processes in the earth mantle, is far from being well understood. At the melting transition the shear modulus of the material vanishes \cite{Born39}. There are several phenomenological criteria predicting when this will happen (see \cite{Born39,Cahn78,Mei+07} for a discussion). Most prominent is the Lindemann criterion \cite{Lindemann10}, stating that melting occurs when the mean square fluctuation of the atoms in a solid is a fraction $c_{\textrm{L}}$ (the  Lindemann number) of the interatomic distance. 
On a more conceptual level it is expected that the proliferation of defects or grain boundaries plays an important role in melting. In two dimensions melting was indeed shown to be a two stage process, driven by unbinding of dislocations and disclinations, respectively \cite{Nelson+79}, while in three dimensions a sudden increase of the vacancy concentration was observed at the  transition \cite{Frenkel39,Frenkel55,Gorecki76,Cahn78}.

In this Letter we address the question of the melting process in a heterogeneous system. Since inhomogeneities and extended defects occur naturally in any practical system, comprehension of the melting transition in such systems is of broad importance. Vortex lattice in superconductors has been extensively utilized as a model system for theoretical and experimental studies of the melting transition \cite{Blatter+94,Zeldov+95}.
Melting in flux line lattices of type-II superconductors is 
characterized by  the dimensionless parameter $\epsilon_{\textrm T}=\lambda/\Lambda_T$. 
 $\lambda$ denotes the  London penetration depth and  $\Lambda_T$ is  a thermal length scale, only related to the temperature $T$ and the flux quantum $\phi_0=hc/(2e)$,  $\Lambda_T={\phi_0^2}/{(16\pi^2 k_BT)}$.
In high-T$_c$ materials with their elevated transition temperatures and  large values of $\lambda$, $\epsilon_{\textrm T}$ can become of order one replacing a  large part of the  mean field phase diagram by the vortex liquid phase \cite{Sudbo+91}.

It was recently found \cite{Segev+11} that by tilting the magnetic field away from the $c$-axis the melting transition in highly anisotropic high-$T_c$ superconductor Bi$_2$Sr$_2$CaCu$_2$O$_{8+\delta}$ (BSCCO) changes from a homogeneous first-order process into a two-step transition with an intermediate lamellar solid-liquid phase that is apparently driven by an intrinsic heterogeneity of the vortex lattice structure. 
In layered materials like BSCCO the vortex lines along the $c$-axis consist of stacks of pancake vortices (PVs) residing in the superconducting layers, whereas the vortex lines parallel to the $ab$ plane take advantage of the normal layers forming Josephson vortices (JVs) \cite{Clem91,Bulaevskii+92a}. When the magnetic field is tilted away from the $c$-axis, chains of higher vortex density were found in Bitter decorations \cite{Bolle+91,Grigorieva+95,Tokunaga03}, scanning Hall probe microscopy \cite{Grigorenko+01,Grigorenko02}, Lorentz microscopy \cite{Matsuda01}, and magneto-optical imaging \cite{Vlasko-Vlasov02,Tokunaga02,Segev11}, and described as crossing lattices of pancake and Josephson vortices (see Fig. 1a) \cite{Huse92,Koshelev93,Koshelev99,Dodgson02a,Koshelev03,Koshelev05,Bending05}. 
The resulting vortex lattice structure displays a very rich phase diagram that depends critically on the strength and direction of the magnetic field ${\bf H}$ and on the ratio of $\epsilon_s/\gamma$, where $\epsilon_\textrm{s}\equiv\lambda/s$ ($\epsilon_s\simeq 130$ in BSCCO), $\gamma\equiv\lambda_c/\lambda$ is the anisotropy parameter ($\gamma\simeq 500$), $\lambda\equiv\lambda_{ab}$ is the in-plane penetration depth, and $s$ is the interlayer spacing \cite {Koshelev05}. The essential element that characterizes the crossing lattices state is that the PV stacks that reside along the JV planes (see Fig. 1a) experience local environment that is substantially different from the bulk PVs thus leading to formation of a heterogeneous system which is expected to undergo a distinctive melting process. Since the degree of heterogeneity can be readily controlled by the direction and magnitude of the magnetic field, the study of this system could provide an important insight into the microscopic mechanisms of melting in heterogeneous systems.

The previous theoretical studies of the melting of the crossing lattices \cite{Huse92,Koshelev93,Koshelev99,Dodgson02a,Koshelev03,Koshelev05} and most of the experimental studies \cite{Schmidt+97,Ooi99,Mirkov01} have investigated the global behavior assuming a single melting transition. The recent magneto-optical imaging of the melting process has shown, however, that the local process in such a heterogeneous system cannot be described by a single transition and that the planes containing the Josephson vortices melt first creating an intermediate periodic solid-liquid state \cite{Segev+11}. The goal of the present work is to provide a theoretical framework describing the melting of a heterogeneous lattice and of the pre-melting of the Josephson planes. Keeping in mind the complicated lattice structure as well as the absence of a comprehensive theory of 3D melting even for simple systems, we will resort to a mean-field-like vacancy theory of melting. In this description the effective cage potential, acting on a given PV due to its interaction with all other vortices, is calculated by replacing the instantaneous positions of the latter by their equilibrium ones. Melting is concluded to occur at temperatures where the PV under consideration can escape from the cage potential via its lowest saddle point, triggering proliferation of vacancy-interstitial pairs. This happens first for PVs in the Josephson planes. The melting temperature determined in this way is in good agreement with the experimental findings.

 \smallskip

{\it Model.---}We follow here the 
 conclusions of \cite{Koshelev05} that the discrete layer structure has strongest influence on the
cores of tilted and Josephson vortices, but that interaction contributions
to the total energy usually can be computed within
continuous approximation. Pancake and JVs in their equilibrium position can be considered to be  threaded on a contour line $\br_i(t)$. The vorticity ${\bomega}$ of the  lattice can  be written as
\begin{equation}
{\bomega}(\br)=\sum_i\int dt({d\br_i(t)}/{dt})\delta\lk\br-\br_i(t)\rk,
\end{equation}
which is related to the magnetic induction $\mathbf B$ by
\begin{equation}
{\mathbf B}+\bnabla\times\sum_\alpha\lambda_\alpha^2(\bnabla\times{\mathbf B})_\alpha\hat r_\alpha=-\phi_0\bomega(\br).
\end{equation}
Here $\lambda_z\equiv\lambda_c$ and $\lambda_x=\lambda_y\equiv\lambda$.  The free energy $\cF$ of vortex systems is   a functional of the vorticity field which includes contributions both from the magnetic and the Josephson interaction \cite{Koshelev05} \begin{equation}\label{eq:F}
 \cF[\bomega]=\!\frac{\phi_0^2}{8\pi}\!\int\frac{d^3k}{(2\pi)^3}
 \frac{|\bomega_{\mathbf k}|^2+k^2(\lambda_{c}^2|\omega_{z{\mathbf k}}|^2+\lambda^2|{\mathbf \bomega}_{\perp{\mathbf k}}|^2)}{(1+\lambda^2k^2)(1+\lambda^2k_z^2+\lambda_c^2{\mathbf k}_\perp^2)}.
\end{equation}
$\bomega_{\mathbf k}$ 
 is the Fourier transform of  $\bomega(\bf r)$. The subscript $\perp$ denotes the projection of a vector onto the $xy$-plane.  
 The actual equilibrium flux lattice structure follows from the minimization of 
 \begin{equation}
 \cF[\bomega]-\int d^3r{\mathbf B}\cdot{\mathbf H}/{4\pi}  \end{equation} 
for a given magnetic field $\mathbf H$.
  Its determination   is in general a  complicated task because of the potentially large number of competing ground states.
   In this paper we will not address this question but instead assume that  the structures found experimentally are  indeed the free energy minima for a given applied magnetic field $\mathbf H$. Further below we will consider the effects of a possible incommensurate superstructure and find that these are weak.
   \begin{figure}[h] \includegraphics[width=8.5cm]{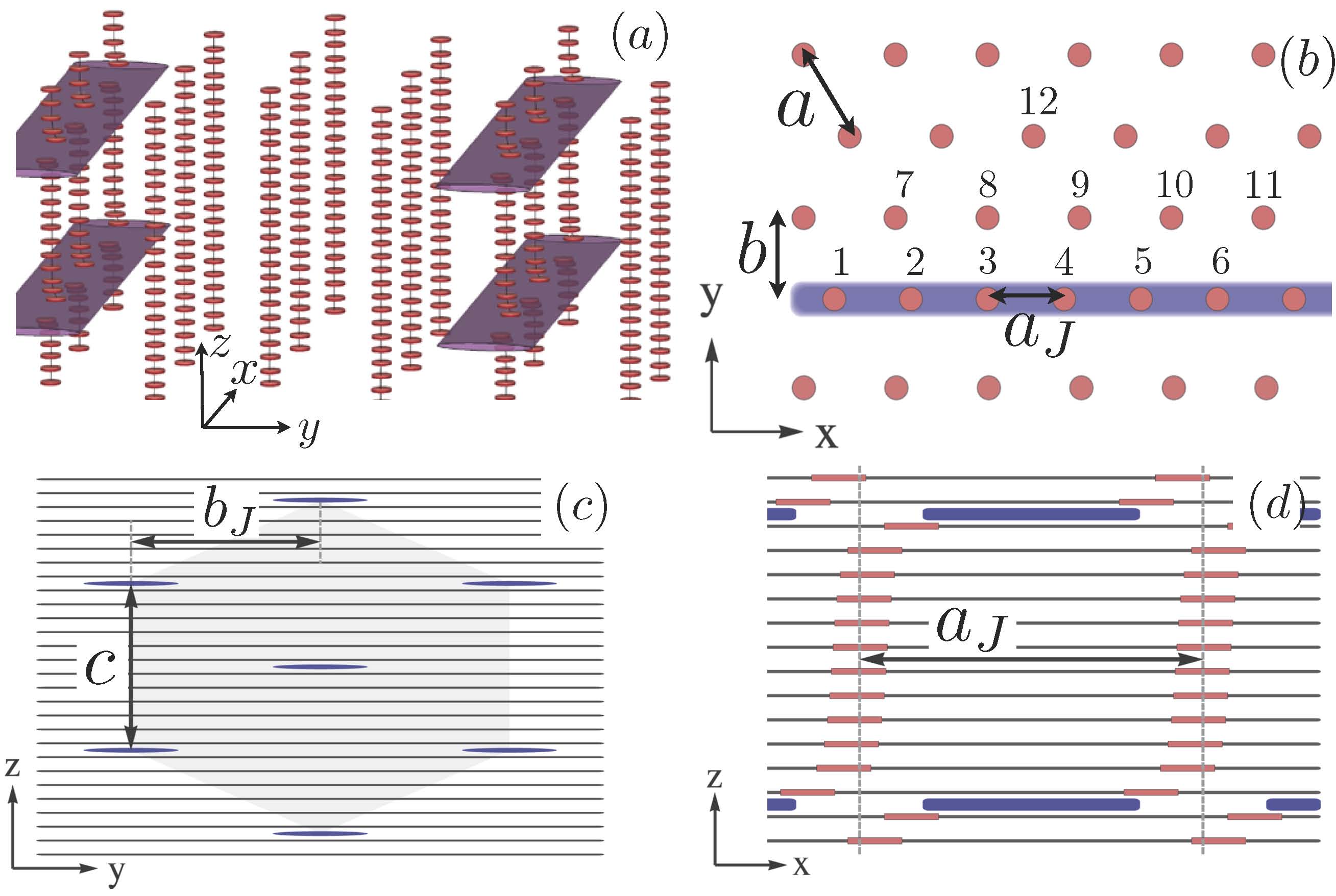} 
    \caption{ (a) Schematic 3D plot of the crossing PV-JV lattice. The JVs are aligned along the $x$ axis and are stacked along $z$ forming ``Josephson planes" along $xz$-planes. (b) Top view ($xy$-plane) of the PVs (red) and JVs (blue). The vortex numbers are used in Figs. 2 and 3. (c) Front view of the JV lattice ($yz$-plane). (d) Side view of PVs and JVs in the Josephson plane ($xz$-plane).
    }
   \label{fig:example}
\end{figure}

 \smallskip   
Once the equilibrium structure is known,  the effective potential for  a pancake vortex displacement $\bu_{i,n}$ in stack $i$ and layer $n$   can be found from  the change of  the  vorticity field $\Delta\bomega(\br)$ in (\ref{eq:F}). Its Fourier transform is 
$\Delta\bomega_{\mathbf k}=\hat zs e^{i{\mathbf k}{\mathbf r}_i}\lK e^{i{\bf k}_\perp {\mathbf u}_{i,n}}-1\rK
 $.  This allows to  calculate the free  energy change 
$\Delta\cF(\bu_{i,n})$ which depends in general on $i$ and $n$.  
The actual vortex structure in the Josephson planes consists of  continuous vortex lines formed of stacks of PVs, connected by JVs, as indicated in Fig. \ref{fig:example}d \cite{Buzdin+02,Koshelev05}. The resulting discontinuity of the JVs in a single plane (and hence also the displacements of the PVs from a straight stack) is of the order $\lambda^2/[(2n-1)\gamma s]$ \cite{Buzdin+02}, which is for our parameter values $\lesssim0.06\, a_J/n$.   $n\ge1$ denotes the number of layers  between the JVs \cite{Buzdin+02}. We will therefore ignore in the following this small effect and assume a simple crossing lattice of straight stacks of PVs and straight JVs \cite{Dodgson02b}. 

 \begin{figure}[h] \includegraphics[width=8.5cm]{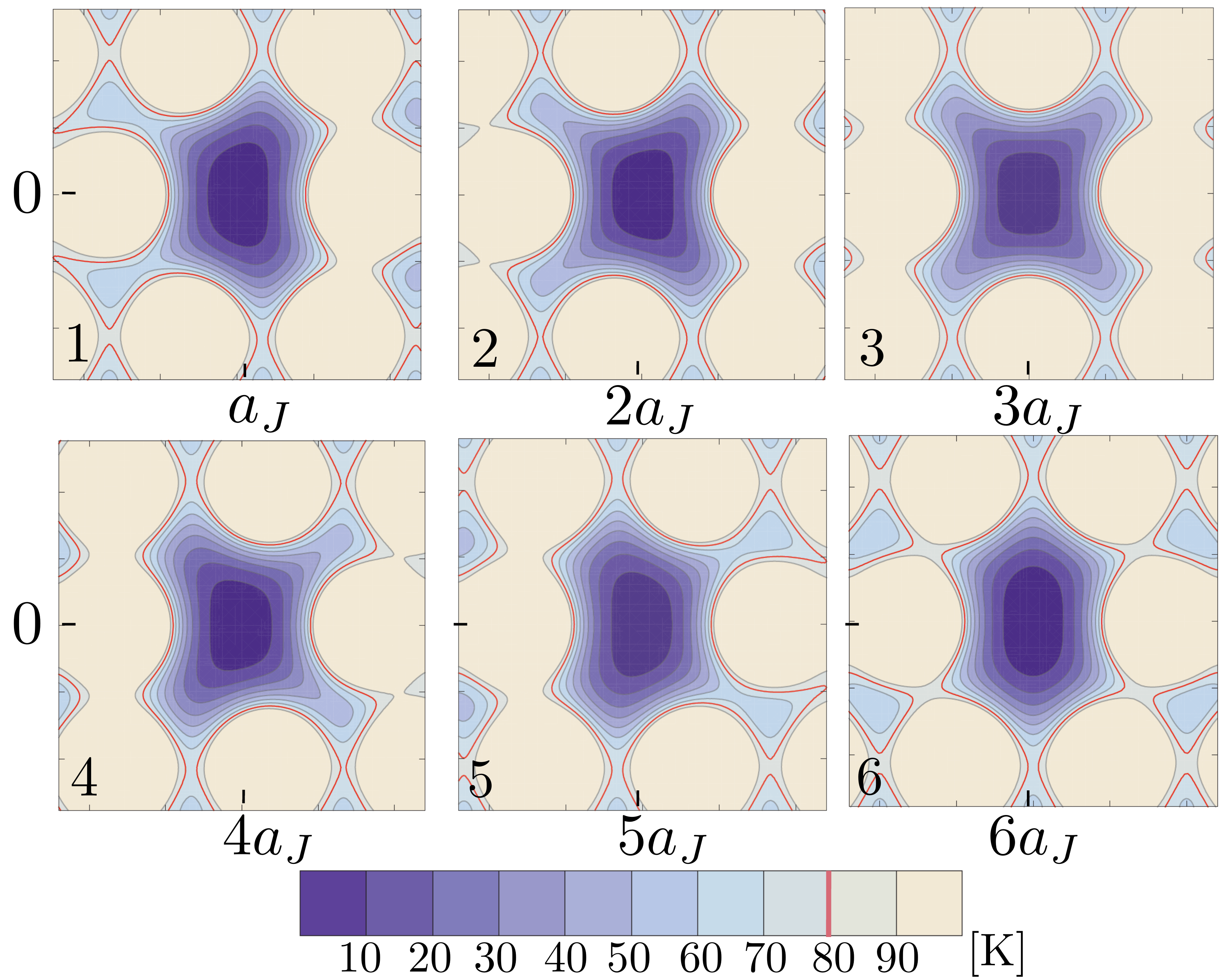}  
 \caption{Set of contour plots of the cage potential $\Delta \mathcal{F}(\mathbf u_i)$ for a single displaced PV in a stack of PVs in the Josephson plane for the different stacks numbered as in Fig. 1b. Only the PV with rest position in the center is assumed to be displaced, keeping all other PVs at their equilibrium positions. With the choice of $a_J/a=5/6$ this sequence is repeated every 6 stacks.} \label{fig:mainplane}
 \end{figure}

Displacing a single  PV from a straight stack generates an antiparallel pair of JVs in the adjacent  non-superconducting layers. Because of their large mutual annihilation, the interaction of this excited JV pair with the static JVs of the crossing lattice will be very small, provided the displaced PV is not too close to the JV.  We will ignore this weak interaction in the following as well. With these approximations   $\hat\omega$ of PV and JVs are now perpendicular to each other, and hence Josephson vortices do not contribute to the effective potential of the displaced pancake vortices once the structure is fixed. Moreover, the effective potential for a single displaced PV does not depend on its position in the stack. This approximation is justified the more the larger the distance between the displaced PV and the JV. 

After summation over the PVs in the  stacks  and suppressing the layer index $n$, the effective potential $\Delta\mathcal{ F}[\bu_i]$ for a single  PV, displaced from a straight stack by $\bu_i$, can be written in the form 
\begin{equation}
{\Delta\mathcal{ F}}=\frac{k_BT}{\epsilon_\textrm{T}\epsilon_\textrm{s}}\sum_{j(\neq i)}\lK K_0\lk{|{\bf R}_{ij}-{\bf u}_i|}/{\lambda}\rk-K_0\lk{ R_{ij}}/{\lambda}\rk\rK.
\end{equation}
Here $K_0(x)$ is the modified Bessel function,  ${\mathbf R}_{ij}$ the position vector connecting  stacks $i$ and  $j$, and $R_{ij}=|{\mathbf R}_{ij}|$.  The remaining sum has to be performed numerically, which requires the knowledge of the flux lattice structure. We will  make  some assumptions for the ground state,  concluded both  from  analytical theory and  experimental findings  \cite{Bolle+91,Koshelev05}:

(i) All vortices are located in planes parallel to the $xz$-plane, assuming  $H_y=0$ (Fig. 1). Planes including crossing PVs and JVs are called Josephson planes. 
 
 (ii) PV stacks outside the Josephson planes are assumed to form a equilateral triangular lattice of spacing $a$ \cite{Grigorieva+95,Koshelev99}. 

(iii) The PV distance in the Josephson planes is denoted by $a_J<a$ and the distance between  Josephson and the adjacent planes is denoted by $b>\sqrt{3}a/2$ (Fig. 1b). $a_J<a$ is the result of the attractive interaction between PVs and JVs. 
  \begin{figure}[h] 
  \includegraphics[width=8.5cm]{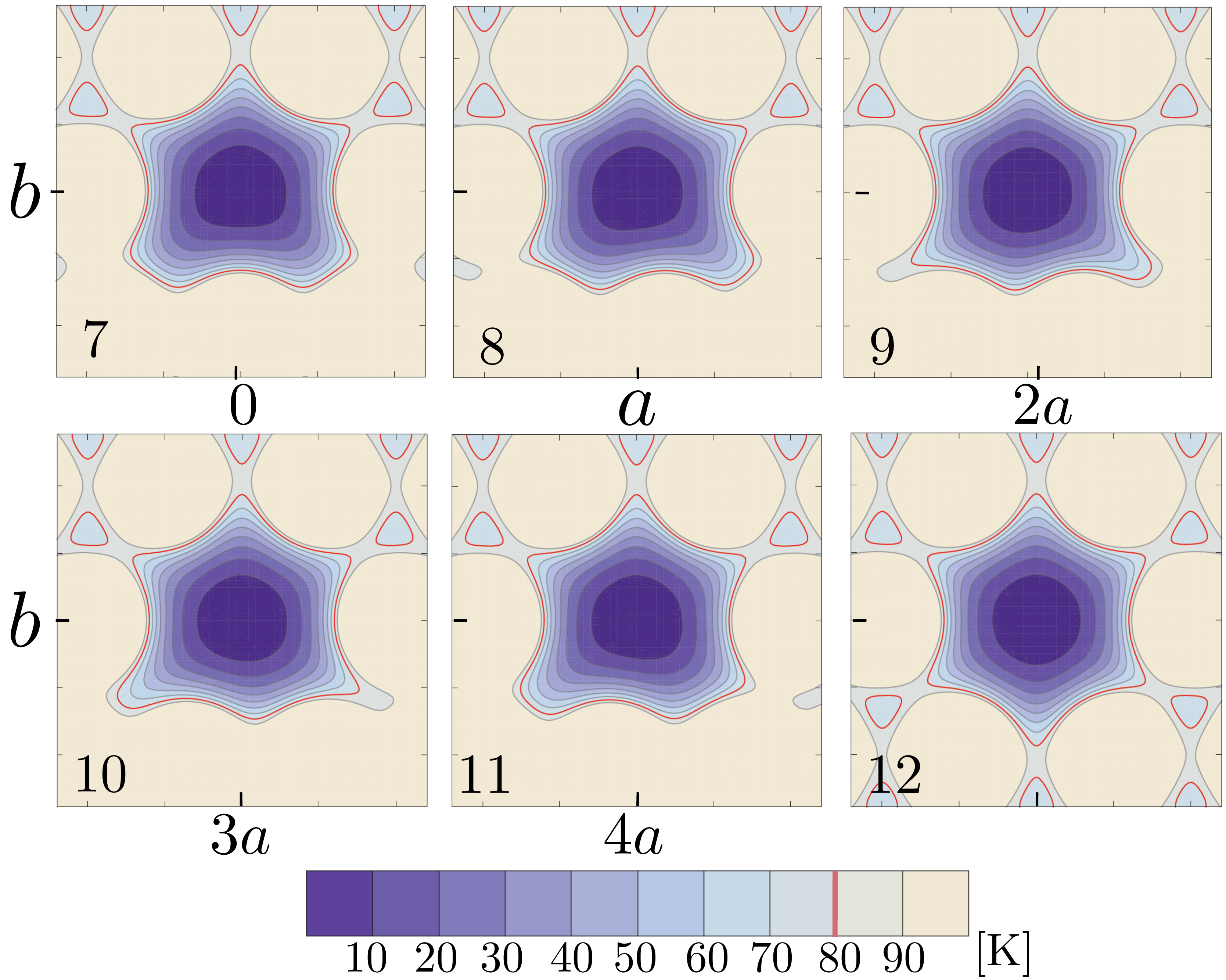}
   \caption{Contour plot of $\Delta \mathcal{F}$ for PVs in the plane adjacent to the Josephson plane (numbered 7 to 11 in Fig. 1b) and in the bulk (number 12). Melting occurs at about $10$ K higher than for PVs in the Josephson plane in Fig. 2.}
   \label{fig:sideplane}
\end{figure}

 (iv) The projection of the JVs onto the $yz$-plane  forms squeezed Abrikosov lattice with the lattice parameter ratio    
$b_J/c=\sqrt{3}\gamma/2$ (Fig. 1c). 

The above conditions can be somewhat relaxed as discussed below. 
 
 \smallskip   
To be specific we used the lattice structure found in   the Bitter decoration experiment \cite{Bolle+91} performed in BSSCO  at $\mathbf H=(30,0,24)$\,Oe where  $a=1,\, a_J=0.833,\, b=0.91,\,c= 0.04,\,b_J =17.32,\,\lambda(T)=\lambda(0)\sqrt{1-T/T_c},\,\lambda=0.2 $, all lengths in $\mu$m.  The value of $H_x$ is slightly larger than that used in  \cite{Segev+11}.
Figure \ref{fig:mainplane} shows the results of the calculation of the free energy change $\Delta{\cal F}$ due to the displacement of the individual PVs in the Josephson plane labeled 1 to 6 in Fig. 1b. Since in the present case $a_J/a=5/6$, the potential shapes recur every six vortices. The contour maps of lines of constant energy are depicted in units of $k_BT$. There are two important observations here. First, although the precise shape of the potential is different for the individual vortices, saddle points with very similar value $T\approx 75$ K appear for all the locations. At this temperature PVs can escape in direction transverse to the Josephson plane, vacancy-interstitial pairs proliferate, and the lattice starts to melt along the Josephson planes. This transverse melting process appears to be uniform for the various PVs along the Josephson plane. The second observation is that in contrast to the transverse direction, the shape of the potential along the Josephson plane is rather position dependent. Vortices that are located at symmetric points that are in registry with the adjacent planes, like vortex 3 and 6, show potential that is quite localized in the $x$ direction. On the other hand, the potentials at asymmetric locations, like vortex 1 and 5, show extended protrusions that indicate enhanced vortex mobility at the less-stable vortex positions along the Josephson planes. This finding can explain the apparent observation of site dependent enhancement of vortex fluctuations along the Josephson planes in Lorentz microscopy studies \cite{Matsuda01} that was discussed in terms of the incommensurability of the vortex chains structure.

The corresponding pictures for PVs in the plane adjacent to the Josephson plane are shown in Fig. 3 along with PV number 12 in the bulk. Despite of the fact that each of the PVs has a different local energy landscape, it is clearly seen that all the PVs can leave their positions only at temperatures of $T_m\simeq 85$ K, $10$ K higher than the PVs in the Josephson planes. The experimental bulk melting temperature is $T_m\approx90$ K.
 A remark is in order: since the vortex cage potential is calculated with all surrounding vortices in their equilibrium positions, proliferation is predominant in the direction perpendicular to the Josephson plane. We expect that in a treatment which allows simultaneous motion of all vortices this prevalence is reduced.

\begin{figure}[h]
 \includegraphics[width=8.5cm]{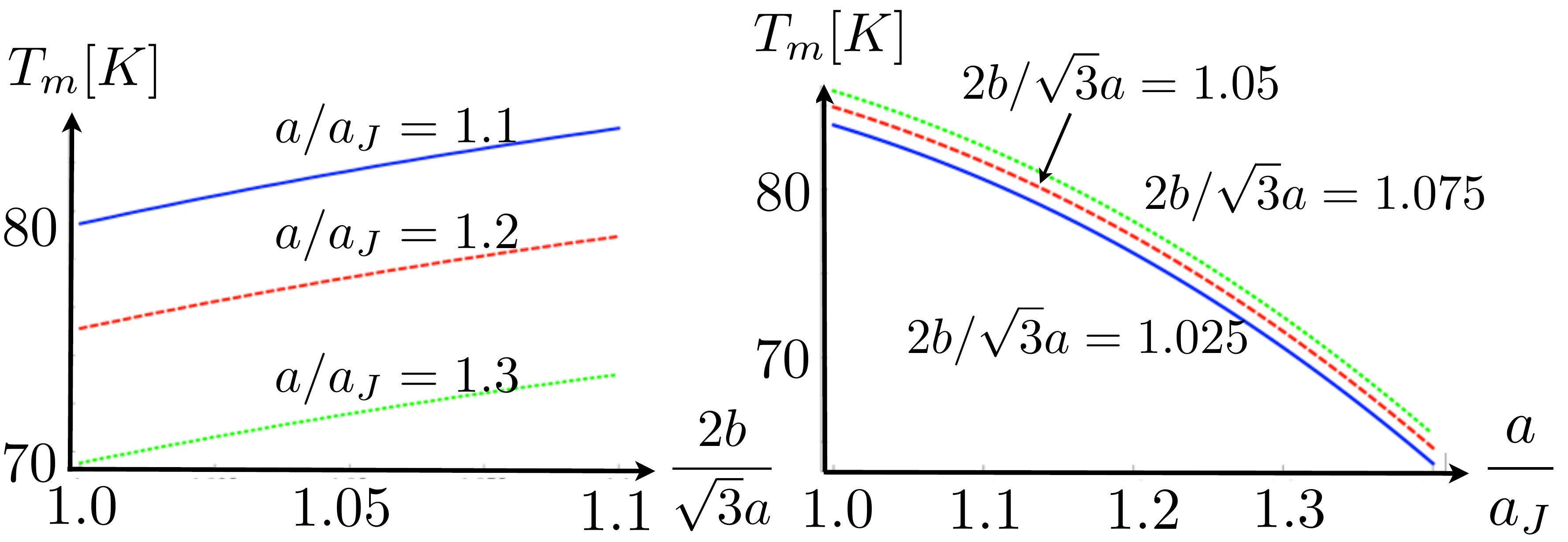} \caption{Melting temperature of PVs along the Josephson planes $T_m$ in crossing lattices state with regular lattice parameter $a=1 \mu$m as a function of $b/a$ (left panel) and of $a/a_J$ (right panel). }\label{fig.4}
 \end{figure}

Figure \ref{fig.4} shows a detailed calculation of the melting temperature $T_m$ of the PVs in the Josephson planes as a function of their intervortex distance $a_J$ and the separation $b$ between the Josephson plane and the adjacent PV plane (see Fig. 1b). $T_m$ is found to drop rapidly with increasing $a/a_J$ and with decreasing $b/a$ giving rise to pre-melting of the Josephson planes that could in principle be as large as $20$ K. In reality, however, $a_J$ and $b$ are not independent parameters and are both determined by $H_x$ and $H_z$. In particular, increasing $H_x$ decreases $a_J$ but increases $b$, so that the two dependences moderate each other to a large extent. We therefore expect a much smaller pre-melting as observed experimentally \cite{Segev+11}. %

\smallskip

So far we assumed equidistant vortex positions in the Josephson plane. This assumption can be somewhat relaxed by allowing a modulation of the  PV spacing, i.e. $a_J\to a_{J}(n)$ where $n$ numbers the intervals. 
The $a_{J}(n)$ follow  from the competition between the PV-JV interaction in the Josephson plane   on one side, which favors a lattice spacing smaller than $a$, and their interaction with  the PVs outside the Josephson plane on the other side, which favor registry with the lattice constant $a$, which we assume to be fixed. 
Such a system can be described by the Frenkel-Kontorowa model \cite{Frenkel+38}, 
\begin{equation}\label{FK-model}
{\cal H}=\sum\limits_n\lK\lk \theta_{n+1}-\theta_n-\delta\rk^2-2\zeta\cos \theta_n\rK.
\end{equation} 
The $\theta_n$ describe the modulation of  the PV  positions,   $x_n=an+a\theta_n/({2\pi})$,  
$\zeta={2\pi^2k_2}/({q^2 k_1})$, and $q=a/\lambda$. 
The misfit parameter   $\delta <0 $, depending on $H_x$, favors a higher PV density in the Josephson planes. It will be determined below from the average vortex spacing $\langle{a_J(n)}\rangle\equiv a_J$.
 The coefficient $k_1$ follows from the vortex interaction as 
$
{k_1}= K_0\lk q\rk+{q^{-1}}K_1\lk q\rk.  
 $
$k_2$ is determined numerically by fitting the actual vortex interaction
of a Josephson plane stack with its neighboring stacks in  the satellite plane to a cosine model, 
taking up to the  fifth next neighbour  into account.
 The ground state configuration follows from the saddle point condition
$\label{saddlepoint}
\theta_{n+1}+\theta_{n-1}-2\theta_n=\zeta\sin \theta_n.
$ 
In the continuum limit, $\theta_n\to \theta(n)$, one arrives at  the rigid pendulum equation $
\theta''(n)=\zeta \sin \theta.
$ 
Its solution can be expressed by elliptic functions which depend on 
the 
constant of integration $\eta$. $\eta$   follows  from the minimization  of 
$K(\eta^2-{2})+{4}{E} -{\pi\delta }{\eta}/\sqrt{\zeta}$.
 Here $K(\eta)$ and $E(\eta)$ are the complete elliptic integrals of first and second kind, respectively. 
For small misfit, $|\delta|<\delta_c=4\sqrt{\zeta}/\pi$, $\theta(n)$ locks-in at a multiple of $2\pi$,  corresponding to $a_J=a$. 
For $|\delta|\gtrsim|\delta_c|$ the solution for $\theta(n)$ is staircase-like with  horizontal terraces at $\theta(n)\approx2\pi p$ ($p\in\mathbb{N}$),  which are connected by steps of width $\sim\zeta^{-1/2}$.  The average PV spacing in the Josephson plane is in general incommensurate with $a$. For a given $a_J$, $\delta$ is determined from the relations
$
a_J=a\lK1-{\sqrt{\zeta}}/({2\eta K})\rK$ and $\eta={4\sqrt{\zeta}E}/({\pi\delta})$.
 With the parameter values given above we get  $\zeta=0. 409$, $|\delta_c|=0.814$, 
  $\eta=-0.876$ and $\delta\approx-1.115$. Thus the system is indeed in the incommensurate phase .   
  The resulting modulation of the vortex distance
$a_J(n)$, however, 
 is less than $6\%$ of the mean vortex distance and hence can be safely ignored for the investigation of the melting transition.

In summary, using a cage potential model we have calculated the melting temperature of the PVs at different locations in the crossing lattices state parametrized by three different lattice constants $a, a_J,b$. Detailed results were obtained for a given PV lattice structure relevant to highly anisotropic superconductor BSCCO. We showed that melting of crossing flux line lattices sets in along the planes containing JVs where the local melting temperature can be substantially lower than in the bulk resulting in a periodic solid-liquid structure as observed experimentally \cite{Segev+11}. We have analyzed the effect of the incommensurate modulation of the PV lattice constant $a_J$ along the Josephson plane and found it to be too weak to have a significant effect on the melting transition. The primary cause of the pre-melting transition is proliferation of vacancy-interstitial PV pairs at reduced temperatures due to the increased PV density along the Josephson planes.

We thank A. Koshelev for useful discussions.
This work was supported by the German-Israeli Research Foundation (GIF) and the DFG (SFB 608).
\bibliography{Vortex_Lattice_Melting-1.bib}
\end{document}